# Voltage-Controlled Nano-Scale Reconfigurable Magnonic Crystal

Qi Wang[1,2,3], Andrii V. Chumak[2], Lichuan Jin[1], Huaiwu Zhang[1], Burkard Hillebrands[2], and Zhiyong Zhong[1]

1. *State Key Laboratory of Electronic Thin Films and Integrated Devices, University of Electronic Science and Technology of China, Chengdu, 610054, China*
2. *Fachbereich Physik and Landesforschungszentrum OPTIMAS, Technische Universität Kaiserslautern, 67663 Kaiserslautern, Germany.*
3. *National Engineering Research Center of Electromagnetic Radiation Control Materials, University of Electronic Science and Technology of China, Chengdu, 610054*

**Abstract**

A nano-scale reconfigurable magnonic crystal is designed using voltage-controlled perpendicular magnetic anisotropy (PMA) in ferromagnetic-dielectric hetero-structures. A periodic array of gate metallic stripes is placed on top of a MgO/Co structure in order to apply a periodic electric field and to modify the PMA in Co. It is numerically demonstrated that the application of the voltage to the gate stripes modifies the spin-wave propagation and leads to the formation of band gaps in the spin-wave spectrum. The band gaps are dynamically controllable, i.e. it is possible to switch band gaps on and off within a few nanoseconds. The width and the center frequency of the band gaps is defined by the applied voltage. At last, it is shown that the application of the voltage to selected, rather than to all gate stripes allows for a pre-defined modification of the band gap spectra. The proposed voltage-controlled reconfigurable magnonic crystal opens a new way to low power consumption magnonic applications.

Spin waves (SWs) are the eigen excitations of a spin system of a magnetic material. They have wavelengths orders of magnitude shorter than the wavelengths of electromagnetic waves of the same frequency, and thus, allow for the design of nano-scale devices[1-4]. Recent experimental and theoretical discoveries have shown that there is much potential for the use of spin waves for transfer[5-7], processing [1-4, 8-13], and (short-time) storage of information[11, 12, 14]. An external control of spin waves, which is of key importance in these prototype devices, was realized using an application of e. g. electric current-induced magnetic field[7-9, 15], spin-polarized current[16], skyrmions[17], another magnons[10], or laser light[18]. The spin-wave control by an electric field opens an access to a new type of magnonic devices with decreased energy consumption[4, 11-13, 19-22]. Here we use electric field-induced perpendicular magnetic anisotropy (PMA) to develop a nano-scale reconfigurable magnonic crystal (MC) that is an universal magnonics component suitable for versatile operations with data[3, 23-25].

Magnonic crystals, which are counterparts of photonic crystals, are magnetic materials periodically-modulated in space [1-3, 10, 14-18, 23-29]. When compared with uniform media, MCs modify the SWs spectra by creating forbidden bands (band gaps) at the Brillouin zone boundaries of the crystal due to Bragg reflection. In recent years, many different kinds of MCs have been reported. They were fabricated by periodic modification of geometrical sizes[26-30], external magnetic field $H_{ext}$ [15, 25], saturation magnetization $M_s$[18, 30], or using periodically varied magnetic materials[31]. However, most of these MCs exhibit static properties defined by the fabrication. Reconfigurable (see review [3]) and fast dynamic magnonic crystals[15, 25], properties of which can be dynamically changed, open additional access for data processing such as reconfigurable filters[8], time reversers[23], and logic devices[25].

Recently, the voltage-controlled perpendicular magnetic anisotropy (PMA) has been intensively studied theoretically as well as experimentally[32-42]. The PMA in ferromagnetic-dielectric heterostructures can be controlled by a voltage applied across the structure due to charge accumulation at the interface between metallic and oxide layers, caused by the resulting electric field. The surface magnetic anisotropy $K_s$ changes nearly linearly with the applied electric field: $\Delta K_s=\beta_s E$, where $\beta_s$ is the magnetoelectric (ME) coefficient that is defined by the composition and the thickness of the ferromagnetic film and the adjacent dielectric layer. This effect has already been utilized in magnetic random-access memory [35,36], for the motion of domain walls [37-40], or for the excitation of ferromagnetic resonance[19] and spin waves[21]. Here we explore the additional application of voltage-controlled PMA, namely to realize reconfigurable, voltage-controlled magnonic crystals.

Micromagnetic simulations were performed using the public Object-Oriented Micromagnetic Framework (OOMMF) [43]. The material parameters used in the simulations are those of Co: the saturation magnetization $M_s=5.8\times10^5$ A/m, the exchange stiffness $A=1.5\times10^{-11}$ J/m, gyromagnetic ratio $\gamma=2.211\times10^5$ m/As, damping constant $\alpha=0.01$. The damping at the ends of the waveguide ($x<30$ nm and $x>1970$ nm) is assumed to be more than 90 times that of other parts to prevent SWs reflection. The cell size used in the simulation is $2\times2\times1.5$ nm$^3$, which is below the exchange length of this material. A sinc field pulse, $h_y(t)=H_0\sin(2\pi\omega_{max}t)/(2\pi\omega_{max}t)$, is applied locally (in volume $2\times40\times1.5$ nm$^3$) at 100 nm from the left edge of the waveguide to excite SWs with a wide frequency range, where $H_0=1000$ Oe and the cutoff frequency $\omega_{max}=200$ GHz. In order to study the SWs in Damon-Eshbach (DE) geometry [44], an external magnetic field $H = 1000$ Oe is applied in $y$ direction, which is large enough to orient the magnetization along the width direction where the wave vector of the SWs is perpendicular to the field along the $x$ direction (see Fig. 1(b)). The propagation of the SWs is determined by collecting the dynamic out-of-plane magnetization $M_z$. The $M_z$ of each cell is collected over a period of $T=10$ ns and recorded with a sampling rate of $T_s=1$ ps, which allows for a frequency resolution of $\Delta f=1/T=0.1$ GHz while the highest resoluble frequency is $f_{max}=1/2T_s=500$ GHz for the excited SWs. The dispersion curve is obtained by performing a two-dimensional (2D) Fast Fourier Transform (FFT) of the temporal $M_z$ in space and time. The transmission characteristic is obtained by integrating the spin wave intensity from 1800 nm to 1900 nm along the $x$-axis.

The structure under investigation is schematically depicted in Fig. 1(a). The MC consists of a 1.5 nm-thick Cobalt (Co) strip grown on the substrate and covered with a MgO layer. This kind of Co/MgO ferromagnetic-dielectric heterostructures have been widely studied[39-42]. Gate metallic strips of $L =10$ nm width are periodically placed on top of the MgO layer with a $P = 30$ nm periodicity (see Fig. 1(b)). This periodic structure is 1.6 µm long, includes 53 gate metallic strips, and is placed in the middle of the 2 µm-long and $w = 40$ nm wide Co strip. The voltage is applied between these gate electrodes and the Co strip in order to control the PMA at the Co/MgO interface.

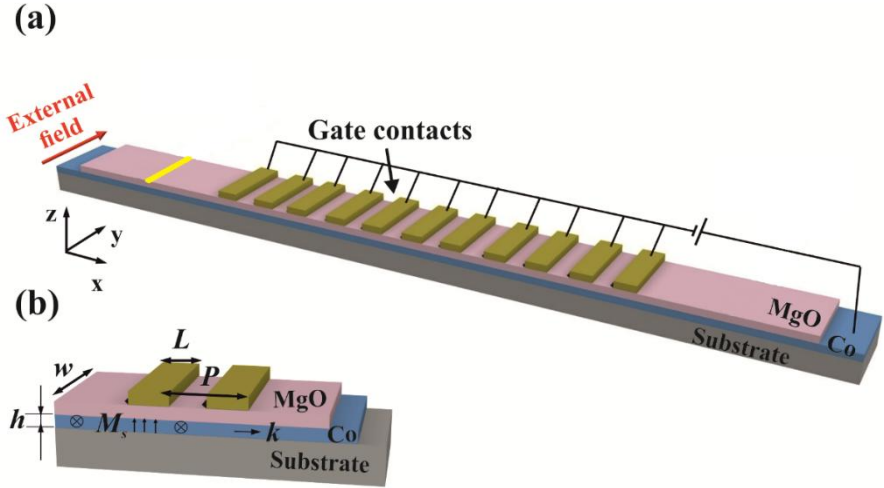

Fig. 1 (a) Schematic view of the voltage-controlled reconfigurable magnonic crystal. (b) Close-up view of the waveguide. An ultrathin Co film of h = 1.5 nm thickness is grown on the substrate and is covered by an insulating MgO film. The gate metallic stripes are periodically placed on the MgO surface. The width and period of the gate metals are 10 nm and 30 nm, respectively. An electric field is applied between the gate strips and the Co layer in order to induce PMA.

Consequently, voltage-induced PMA induces an effective internal magnetic field into the Co film. The ground state of the ultrathin Co film depends on its thickness. The static magnetization of the film tends to be aligned along the out-of-plane axis when the film's thickness is smaller than a critical value (typically about 0.5 - 1.0 nm). In contrast, the magnetization lies in the plane when the thickness is above this critical value. In this work, we consider the in-plane magnetization ground state, but since the PMA is a pure interface effect [32,33,41,42], we choose the thickness of Co strip to be as small as possible $h$ = 1.5 nm. We assume that the PMA value varies linearly with the electric field $E$, i.e., $\Delta K_s = \beta_s E$, where $\beta_s$ is the magnetoelectric coefficient. Values of the ME coefficient, ranging from 20 fJ/Vm to 290 fJ/Vm, have been reported in previous studies[21,33,36,39]. In our simulations, an ME coefficient of $\beta_s$ = 100 fJ/Vm is used. Voltage-controlled PMA modulates the local perpendicular component of the anisotropy field $H_z$ under the gate metallic strips and induces a periodic spatial distribution of the internal field and the component of the saturation magnetization $M_z$. The periodic variation of the internal field, consequently, causes partial or total Bragg reflection of spin waves propagating through the MC.

The instant application of the voltage to the gate strips switches the z-component of the magnetization from the uniform to the periodic state within a few nanoseconds as is shown in

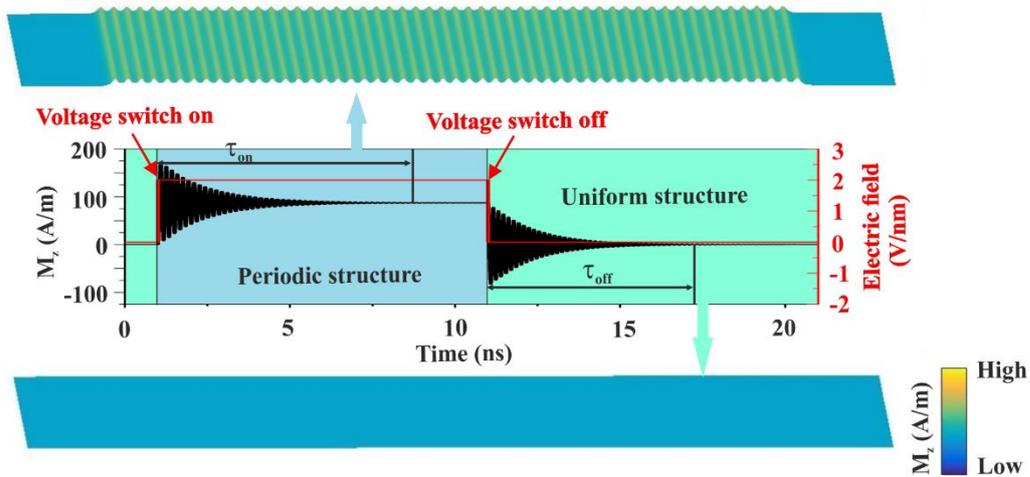

*Fig. 2 The central panel shows the time evolution of the applied electric field (red line) and PMA-induced out-of-plane magnetization $M_z$ (black line). The transient processes are visible after the electric field is applied or switched off. The top and the bottom panels show the spatial distribution of the magnetization for the periodic and the uniform structure, respectively.*

Fig. 2 (see Movie 1 'the dynamic switch process' in the Supplement). The time evolution of the electric field (red line) and $M_z$ (black line) averaged over the whole volume of the sample is shown in Fig. 2. Starting from the static uniform state, an electric field $E$ = 2 V/nm is applied to the gate strips for the time period of 1 ns to 11 ns. The magnetization $M_z$ shows a sharp increase and subsequent oscillatory decay before a stable periodic structure is formed (see the inset in Fig. 2). The duration of the transient process is around 9 ns in our simulations for an electric field of 2 V/nm. After the stationary state was kept for an about 1 ns, the electric field is switched off for the next 10 ns. The periodically-varied magnetization starts to oscillate and gradually relaxes to a uniform state after a transient time of about 6 ns. Thus, it is shown that the properties of the magnonic crystal can be dynamically controlled by an externally applied voltage, and the switching time between the different states is less than 10 ns. The switching time decreases with the decrease in the electric field (see Supplement). Please note that in our simulations the spin-wave propagation time through the periodic structure is approximately 0.8 ns and is, thus, smaller than the on/off switching times. Therefore, this magnonic crystal is termed reconfigurable rather than dynamic. In the latter case, the MC properties are varied faster than the spin-wave propagation time[23]. Nevertheless, the concept presented here can also be used for the realization of dynamic MCs after optimization of the on/off switching times, the group velocity of the spin waves, or the lengths of the periodic structure.

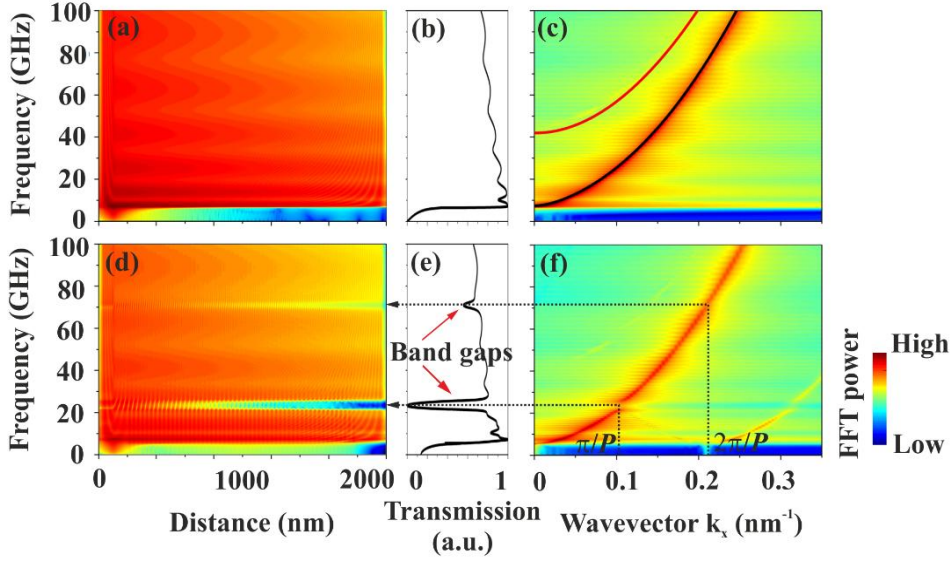

*Fig. 3 The top panels (a)-(c) show spin-wave propagation through the uniform spin-wave waveguide without a voltage applied to the gate stripes. The bottom panels (d)-(f) show spin-wave transmission through the magnonic crystals when a voltage corresponding to E=2V/nm value is applied to the gate stripes. (a), (d) Frequency spectra obtained from a FFT of the temporal evolution of $M_z$. (b), (e) The SW transmission characteristics obtained by integrating the SW intensity over the waveguide area from 1800 nm to 1900 nm along the x axis. (average of all the points along the width direction) (c), (f) The dispersion curves obtained by a 2D FFT of temporal $M_z$ in space and time. The black and the red line in (c) are obtained by analytical calculation of the first and the third width modes, respectively. The black dotted lines in (f) denote the positions of the band gaps and the corresponding wave vectors $k_x$.*

In order to excite spin waves in a wide frequency range from 0 to 200 GHz, a sinc field pulse $h_y(t)=H_0\sin(2\pi\omega_{max}t)/(2\pi\omega_{max}t)$, is applied locally (in the volume $2\times40\times1.5$ nm$^3$) at the 100 nm position from the left edge of the waveguide. In our simulations, magnetic field is $H_0$=1000 Oe and the cutoff frequency $\omega_{max}$=200 GHz. The numerically calculated SW frequency spectra, transmission characteristics, and dispersion curves are displayed in Fig. 3. The top and the bottom panels indicate the characteristic spin-wave propagations in the waveguide with and without an applied voltage in the static regime, respectively. The SW frequency spectra in terms of frequency $f$ versus propagation distance $x$ are obtained by a FFT of $M_z$ on the time scale. The SW transmission characteristics are obtained by the integration of the SW intensity in space over the distance from 1800 nm to 1900 nm along the x-axis. The dispersion curves in terms of frequency f versus wavevector kx are obtained by a 2D FFT of $M_z$ in both space and time (see Supplement).

Low values of the magnetization precession amplitude are shown in Fig. 3 in blue, while high values of the precession amplitude, shown in orange/red. When no voltage is applied to the structure, there are no band gaps observed in the allowed spin-wave band above 6.8 GHz - see Fig. 3 (a-c). Fig. 3 (b) shows that the maximum spin wave intensity corresponds to an FMR frequency of around 6.8 GHz. The intensity decays with an increase in the frequency due to an increase in the damping for high-frequency SWs and a decrease in the excitation efficiency. The SW intensity fluctuations for waves of different frequencies, observed in Fig. 3(b), are due to the interference between the fundamental (black line in Fig. 3 (c)) and 3$^{rd}$ order width modes (red line) as well as between the forward SWs and the SWs reflected from the edge of the structure[45]. The reflected SWs have not been completely suppressed by the absorbing boundary conditions and can be seen in the form of standing waves close to the right end of the waveguide (Fig. 3 (a) and (d)).

The characteristics of propagating SWs are modified when a voltage is applied to the periodic gate strips. Figure 3(d) clearly reveals the formation of the allowed and the forbidden bands in the SW transmission. Two distinct band gaps appear near 23 GHz and 70 GHz with widths of 4.1 GHz and 2.6 GHz, respectively. By comparing the SW transmission characteristic with the SW dispersion curve, one can see that the center frequencies of the band gaps are correlated to the first and the second Brillouin zone boundaries of the MC $k_x = \pi/P = 0.105$ rad/nm and $2\pi/P = 0.21$ rad/nm, respectively. The occurrence of the band gaps at the Brillouin zone boundaries is due to Bragg reflection of the propagating SWs from the voltage-induced, periodically-varied magnetization.

Figure 3 demonstrates one of the key points of this paper: voltage-controlled PMA can induce SW band gaps in a uniform nanostripe waveguide, opening the path to low loss, low power, and voltage controlled band gaps. Further studies show that the width and the center frequency of the band gaps can be dynamically controlled by the applied voltage. Figure 4(a) shows the periodic spatial magnetization distribution for different voltages applied to the gate strips. One can see that the application of a voltage to the gate strips results in the appearance of a pronounced first band gap at around 23.3 GHz. This band gap is already noticeable for a voltage corresponding to an electric field as small as E = 0.5 V/nm (not shown here). The depth of the 1st band gap increases with increasing voltage and reaches its lowest depth for an applied voltage corresponding to E = 2 V/nm. A further increase in the voltage only results in a broadening of the band gap (see Fig. 4(a)). This is a known behavior and was also observed in many magnonic crystals of different types [24-28]. The second band gap is identifiable when a voltage corresponding to an electrical field of E ≥ 1 V/nm is applied.

Figure 4 (b) shows the width (black line) and center frequency (red line) of the first band gap as a function of the applied electric field. The band-gap width increases linearly with an increase in the field and is determined by the amplitude of the magnetization variation which is created by the voltage-controlled PMA. The center frequencies of the band gap are determined by the periodicity $P$ and by the spin-wave dispersion relation. Although the periodicity of the magnonic crystal does not depend on the applied voltage, the spin-wave dispersion depends on the voltage magnitude since it defines the effective magnetization. Spin-wave dispersion might be defined for the case of averaged magnetization[25,26] and shifts towards smaller frequencies with a decrease in

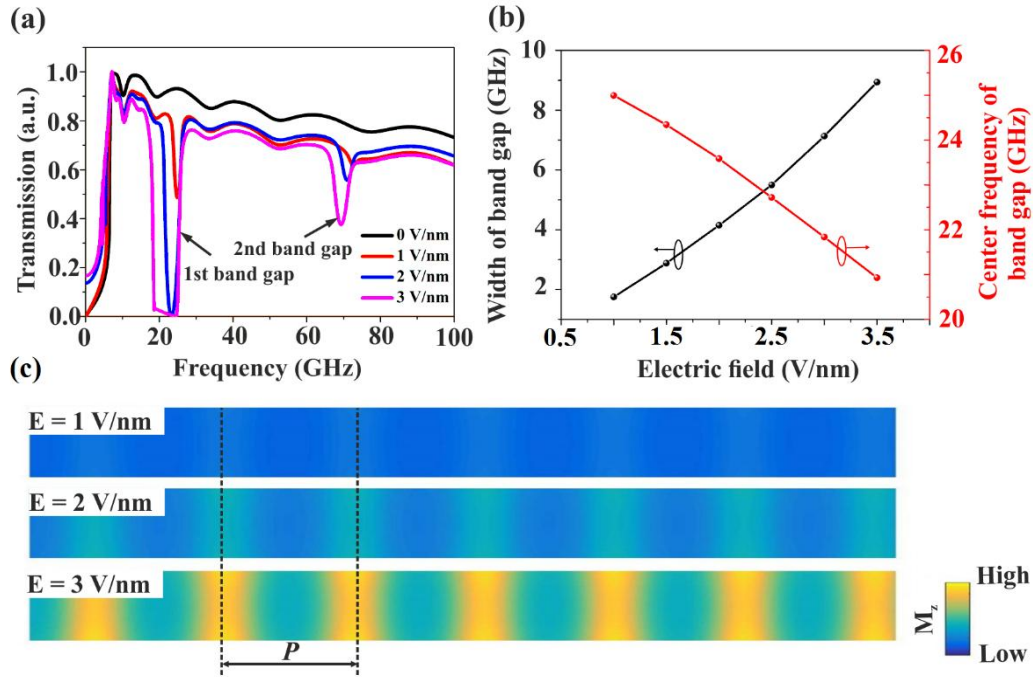

*Fig. 4 (a) Transmission characteristics for various electric fields. (b)The width (black line) and center frequency (red line) of the first band gap versus electric field. (c) Snapshot of the spatial magnetization distribution for different applied voltages.*

the saturation magnetization periodically in space. Similarly, in our case, the voltage increase causes an increase in the out-of-plane magnetization, a decrease in the in-plane magnetization, and, consequently, the shift of the dispersion curve towards lower frequencies. Thus, the center frequencies of the band gaps decrease with an increase in voltage as is shown in Fig. 4(b).

The results presented above show that spin-wave propagation can be dynamically controlled using voltage-induced PMA. However, an application of the voltage to selected gate strips only, opens an additional degree of freedom for the control of spin-wave properties. A simple example of such an application is depicted in Fig. 5. The top panel of Fig. 5(c) shows the top view of the waveguide, where the periodic yellow rectangles represent the metal gate strips and pink represents the waveguide. The width and periodicity of the gate stripes are 10 nm and 20 nm, respectively. All the metal gates are periodically encoded by the numbers "1", "2", and "3" as shown in Fig. 5(c). The voltage is either applied to all gate strips (gates 1, 2, 3-black line), to only strips number 1 and 2 (red line), or only to number 1 strips (blue line). The spatial distributions of the $z$-component of magnetization for these three particular cases are shown in Fig. 5 (c) in the bottom panels. The periodicities of the magnonic crystals are changed by applying the voltage in different manners and, as a result, the spin-wave spectra are modified in different ways. For the first case, when the voltage is applied to all the gate strips, only one band gap is observed at a frequency of about 42.2 GHz – see black line in Fig. 5(a). In the second and third cases, we observe two additional band gaps at frequencies of around 23.5 GHz and 10.6 GHz, respectively. To interpret this result, we perform a FFT of the magnetization $M_z$ along the $x$ direction and combine the dispersion curve obtained by simulation (color map) and theory (black and red lines, for details see the calculations in the Supplement). The FFT provides the magnetization distribution in reciprocal space as shown in Fig. 5(d). Combining the dispersion curve, the existence/absence of each band gap is clearly visible. We can dynamically apply the voltage to different gates to obtain programmable transmission characteristics for the propagation of the SWs. The above-mentioned voltage-controlled reconfigurable MC is a simple example. Many other structures can be designed to realize more complex functions.

In summary, we have demonstrated a reconfigurable magnonic crystal based on the use of voltage-controlled perpendicular magnetic anisotropy. The voltage-controlled PMA induces a spatial variation of the internal field along the waveguide resulting in pronounced band gaps in the

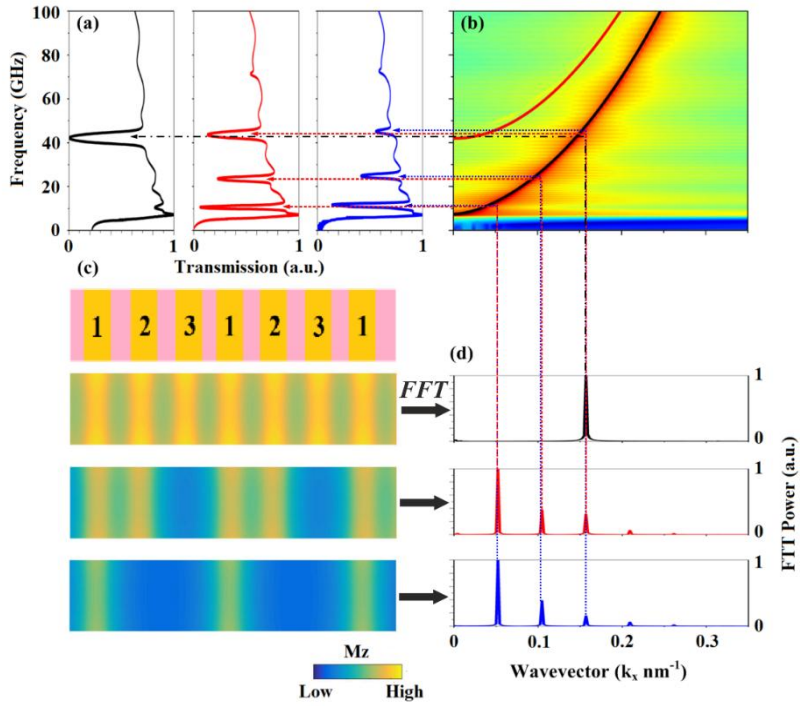

*Fig. 5 (a) Transmission characteristic of different structures. (b) The dispersion curve obtained by micromagnetic simulation (color map) and analytic theory for first and third width modes (black and red line, respectively). (c) The first line is a top view of the waveguide while the others are the spatial magnetization distribution for different structures of the voltage gates. (d) The Fourier transform of the magnetization along the x direction for different structures of voltage gates shown in panel (c).*

spin-wave spectra. The band gaps are dynamically controllable, i.e., it is possible to switch the band gaps on and off within 10 nanoseconds. The width and the center frequency of the band gaps can be tuned by the applied voltage. Finally, we designed a reconfigurable magnonic crystal whose band gaps can be dynamically controlled by applying a voltage to different metallic gates. Our proposal opens the path to low loss, low power voltage-controlled reconfigurable magnonic crystals that can be used in future nano-sized magnon spintronics devices.

Acknowledgements:

This paper is supported by National Nature Science Foundation of China under Grant Nos. 61571079, 61271037 and by Open Foundation of National Engineering Research Center of

Electromagnetic Radiation Control Materials (ZYGX2014K003-9). The project was financially supported partially by the European Union via the ERC Starting Grant 678309 MagnonCircuits and partially by the DFG via SFB/TRR 173: SPIN+X.